\documentclass[twocolumn,pre,superscriptaddress]{revtex4}

\usepackage{amsmath}
\usepackage{amssymb}
\usepackage{calrsfs}

\usepackage{graphicx}

\newcommand{\dd}{\mathrm{d}}
\newcommand{\e}{\mathrm{e}}

\newcommand{\defn}{\textit}
\renewcommand{\vec}{\mathbf}

\newcommand{\Ord}{\mathrm{O}}
\newcommand{\E}{\mathrm{E}}

\begin{document}

\title{The small-world effect is a modern phenomenon}
\author{Seth A. Marvel}
\affiliation{Department of Mathematics, University of Michigan, Ann Arbor,
  MI 48109, U.S.A.}
\author{Travis Martin}
\affiliation{Department of Electrical Engineering and Computer Science,
  University of Michigan, Ann Arbor, MI 48109, U.S.A.}
\author{Charles R. Doering}
\affiliation{Department of Mathematics, University of Michigan, Ann Arbor,
  MI 48109, U.S.A.}
\affiliation{Department of Physics and Center for the Study of Complex
Systems, University of Michigan, Ann Arbor, MI 48109, U.S.A.}
\author{David Lusseau}
\affiliation{Institute of Biological and Environmental Sciences,
  University of Aberdeen, Aberdeen, U.K.}
\author{M. E. J. Newman}
\affiliation{Department of Physics and Center for the Study of Complex
Systems, University of Michigan, Ann Arbor, MI 48109, U.S.A.}
\begin{abstract}
  The ``small-world effect'' is the observation that one can find a short
  chain of acquaintances, often of no more than a handful of individuals,
  connecting almost any two people on the planet.  It is often expressed in
  the language of networks, where it is equivalent to the statement that
  most pairs of individuals are connected by a short path through the
  acquaintance network.  Although the small-world effect is
  well-established empirically for contemporary social networks, we argue
  here that it is a relatively recent phenomenon, arising only in the last
  few hundred years: for most of mankind's tenure on Earth the social world
  was large, with most pairs of individuals connected by relatively long
  chains of acquaintances, if at all.  Our conclusions are based on
  observations about the spread of diseases, which travel over contact
  networks between individuals and whose dynamics can give us clues to the
  structure of those networks even when direct network measurements are not
  available.  As an example we consider the spread of the Black Death in
  14th-century Europe, which is known to have traveled across the continent
  in well-defined waves of infection over the course of several years.
  Using established epidemiological models, we show that such wave-like
  behavior can occur only if contacts between individuals living far apart
  are exponentially rare.  We further show that if long-distance contacts
  are exponentially rare, then the shortest chain of contacts between
  distant individuals is on average a long one.  The observation of the
  wave-like spread of a disease like the Black Death thus implies a network
  without the small-world effect.
\end{abstract}
\maketitle

\section{Introduction}
The \defn{small-world effect} is the observation that it is possible to
connect almost any two members of the world population by a short chain of
acquaintances.  In network terms, the average length of the shortest path
between two nodes in a social network is small.  In practice, ``small''
usually means less than about a dozen steps, even when the network has
billions of members.  Scientific studies of the small-world effect began in
the 1950s with mathematical work by Pool and Kochen~\cite{Pool78}, which
inspired a now-famous series of experiments by Milgram and
co-workers~\cite{Milgram67,TM69}.  Recent years have seen a resurgence of
interest in the topic following publication of an influential paper by
Watts and Strogatz~\cite{Watts98}, and many experiments have been performed
directly confirming the existence of the small-world effect by explicitly
measuring path lengths in
networks~\cite{Albert99,Newman01a,Dodds03,LH08,Backstrom12}.

Given the extensive documentation of the small-world effect, one might
consider it to be an established fact.  It would not be unreasonable to
assume, based on the published literature, that essentially all social
networks exhibit the effect.  Here we argue, however, that this is not the
case.  In particular, we present historical observations and mathematical
results that together suggest that path lengths in social networks used to
be much longer than they are today.  The small-world effect is, we contend,
a modern phenomenon.

Explicit studies of the structure of social networks go back only about a
century---the work of psychologist Jacob Moreno~\cite{Moreno34} in the
1930s is considered one of the earliest examples---so we have no direct
measurements of pre-industrial networks.  Instead, therefore, our
conclusions in this paper are based on historical patterns of the spread of
disease.  Many diseases are passed between individuals via close contact,
be it through shared air, animal vectors, sexual contact, or other means of
transmission.  The pattern of contacts forms a network whose structure in
turn dictates the pattern of infection and hence observations of diseases
can give us hints about network structure.  The physical contact networks
governing disease are not in general the same as acquaintance networks or
social contact networks.  But, to the extent that most social exchange
among humans before the modern era took place via face-to-face interaction,
the network of social contacts was, to a good approximation, a subset of
the network of physical contacts, and this observation allows us to draw
conclusions about social networks as well.

Like other networks, physical contact networks in the modern world appear
to show the small-world effect.  Work on human mobility has shown that the
lengths of trips people take follow a power-law distribution over a wide
range of scales from tens to thousands of
kilometers~\cite{Gonzalez08,Noulas12}, and theoretical work by
Kleinberg~\cite{Kleinberg00} suggests that such a power-law distribution
of connections implies the small-world effect.  Moreover, new strains of
pathogens such as influenza are observed to travel around the globe
rapidly, appearing in distant locations almost
simultaneously~\cite{Earn02}.  One possible cause of such rapid spread is
the presence of short chains of physical contacts linking individuals in
distant parts of the world.

Historically, however, diseases have spread much slower, which raises the
possibility that short contact chains spanning long distances did not
exist, or were much rarer, before the modern era.  But the mere absence of
rapid disease spreading is not, on its own, conclusive.  In their work on
the small-world effect, Watts and Strogatz~\cite{Watts98} showed that only
a vanishing fraction of long-distance trips are necessary to achieve the
effect, in which case it is conceivable that the ancient world was small.
Some long-range connections have existed as far back as reliable records of
civilization extend, being vital to the communication and intelligence
gathering networks of early empires and a necessary part of
intercontinental trade~\cite{Eliot55}.  The question we need to answer,
therefore, is whether there were enough of these long-range connections to
produce a small-world effect in historical contact networks.  In this
paper, we argue that there were not.

\section{Outline of argument}
\begin{figure}
\begin{center}
\includegraphics[width=\columnwidth]{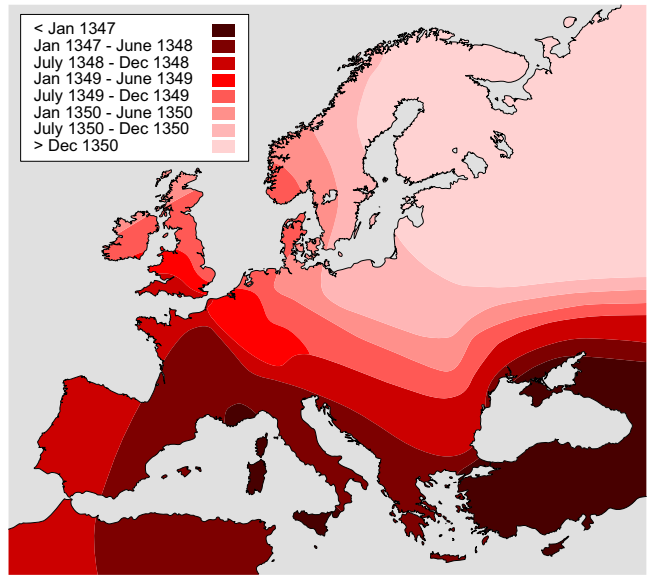}
\end{center}
\caption{The spread of the Black Death across Europe in the 14th century,
  after Sherman and Salisbury~\cite{SS08}.  Observe that the disease
  advanced as a wave of infection across the continent at a more or less
  constant speed for over three years.\label{fig:BlackDeathMap}}
\end{figure}

One can make a heuristic argument for the historical absence of the
small-world effect as follows.  The small-world effect is usually defined
formally as the observation that the typical network path length between
individuals in a population increases no faster than logarithmically with
population size.  Conversely, this implies that the average number of
individuals a given distance away from a randomly chosen starting point in
the network increases exponentially with distance.  And this in turn
suggests that a disease spreading outward from an initial carrier will
infect an exponentially growing number of people as time passes until the
infection saturates the population.  Such exponential growth has been
observed in modern-day disease outbreaks~\cite{CNB07}.  Many historical
outbreaks, on the other hand, show a convincingly non-exponential pattern
of spread---see Fig.~\ref{fig:BlackDeathMap} for an example---which
suggests that the small-world effect was not present at the time these
epidemics occurred.  While this argument is intuitive, however, it is
difficult to make rigorous, so in this paper we take a different approach,
based not on temporal disease patterns but on spatial ones.  In outline,
our argument is as follows.

Our key empirical observation is that, while modern-day epidemics spread
easily and rapidly across vast distances, the same is not true of
historical ones.  Historical epidemics often advanced across the landscape
in a measured, wave-like fashion.  We will focus especially on one
particular example, the 14th-century pandemic known as the Black Death
(Fig.~\ref{fig:BlackDeathMap}).  An unusually destructive outbreak, the
Black Death is thought to have begun in China and then spread along the
Silk Road to the Levant.  From there it was carried on trade ships along
the Mediterranean and, starting in 1347, spread northward across the
European continent, reaching France and Austria in 1348, Germany in 1349,
and Scandinavia and Russia in 1350~\cite{Christakos07}.  As
Fig.~\ref{fig:BlackDeathMap} illustrates, the epidemic displayed a
well-defined wave-front of infection that traveled across the landscape at a
speed of about 800 kilometers per year.

\begin{figure}[t]
\begin{center}
\includegraphics[width=7cm]{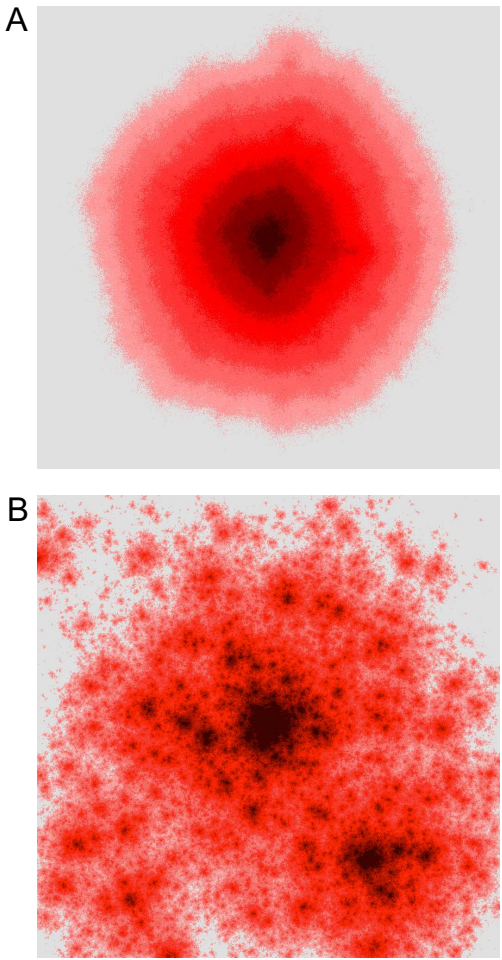}
\end{center}
\caption{Simulations of the spatial spread of an susceptible--infectious
  (SI) epidemic on a $1000\times1000$ grid according to the model in this
  paper.  Each simulation was initiated with a single carrier in the center
  of the system and colors represent the time at which each node contracted
  the disease.  The contact kernel for (A) is an exponential $\alpha(r) =
  a\e^{-br}$.  For (B) it is a power-law $\alpha(r) = r^{-\gamma}$ with $r$
  restricted to be greater than some minimum value~$r_\textrm{min}$ to
  eliminate infinities.  As the figure shows, the exponential kernel gives
  rise to a distinct wave-front of disease propagating out from the center
  while the power-law kernel produces a more nonlocal spread of disease,
  with many satellite outbreaks and no clear
  wave-front.\label{fig:TwoSimulations}}
\end{figure}

Now consider Fig.~\ref{fig:TwoSimulations}, which shows the results of two
different computer simulations of the spread of a disease through a
population.  The population density is the same in both simulations, as is
the average probability of disease transmission between individuals, yet
the two simulations look very different.  In the first simulation (panel~A)
the disease spreads outward in a circular fashion from an initial seed in
the center, creating a clear wave-like front.  In the second (panel~B) the
spread is irregular, with many different centers of infection and no clear
leading edge, even though this simulation also was started from a single
central seed.  The crucial difference between the two simulations lies in
the probability of disease transmission between individuals as a function
of distance.  In the first simulation, probability of transmission falls
off exponentially with distance (i.e., quite rapidly), whereas in the
second it falls off more slowly, following a power law, with a fat tail of
occasional long-distance transmission events that are responsible for the
satellite outbreaks away from the center of the figure.

It turns out that the behavior of the disease in these two cases is not
merely quantitatively different, but also qualitatively so.  No matter how
long the first simulation is allowed to continue, it will always display a
distinct wave-front (provided we give the disease an arbitrarily large
space to expand into).  By contrast, the irregular nature of the second
simulation only becomes more dramatic as time goes by, and eventually all
similarity to wave-like spread is lost.  While the results of
Fig.~\ref{fig:TwoSimulations} are numerical only, we will demonstrate
these conclusions analytically in the following sections.

We will show using a well-established model of the spreading process that
an epidemic can exhibit wave-like spread at constant speed only if the
probability of disease-causing contact between pairs of individuals falls
off exponentially, or faster, with the distance between them.  We then
employ this result as the input to an analysis of the network of physical
contacts between individuals.  We show that if indeed probability of
contact falls off exponentially with distance, then typical path lengths in
the contact network are long and the network does not display the
small-world effect.  Furthermore, if the network of \emph{social} contacts
is, as we have said, a subset of the network of physical contacts, then the
social contact network is also not a small world, since the social network
is formed by removing edges from the physical network, which can only make
paths longer, not shorter.

Given the observation of wave-like spread of the Black Death, we therefore
conclude, subject to the assumptions made in the analysis, that the social
contact network within Europe in the 14th century did not display the
small-world effect.

\section{Epidemic wave-fronts imply rare long-range contacts}
In this and the following section, we present detailed arguments supporting
the claims made above.  Our arguments are based on mathematical modeling of
the geographic spread of disease.  We consider a model in which a disease
spreads through a population of individuals each of whom lives at some
particular geographic location.  Since the transmission of disease between
individuals normally requires close physical proximity, it can only take
place if one individual travels from his or her location to the location of
another, or if both travel to the same third point.  Our model uses a
\defn{contact kernel} to specify the average frequency of such contacts as
a function of where the two individuals live.  For simplicity we take this
kernel to be isotropic (the frequency of meetings is independent of compass
direction) and translation invariant (the same in all parts of the world),
so that it is a function only of the distance~$r$ between individuals.  We
also assume that average frequency of contact falls off with distance
beyond some point, so the kernel is nonincreasing for all $r$ greater than
some nonnegative constant~$R$.

As an illustration of a contact kernel, consider again the Black Death.
The etiology of the Black Death has been the subject of some debate, but
the current consensus, based in part on DNA evidence from mass graves, is
that it was an outbreak of bubonic plague, which is caused by the bacterium
\textit{Y.~pestis} and communicated both directly from person to person and
by rats and fleas~\cite{Haensch10}.  Neither fleas nor rats travel long
distances, however, other than when they are carried by human
transportation, so the contact network over which the disease is
transmitted consists of primarily local contacts plus longer range contacts
that mirror those of the human population~\cite{Gonzalez08}.  The contact
kernel combines these different transmission vectors and transportation
processes into a single quantity that we take as an input to our model.

We also need to model the dynamics of disease spreading.  There are a
number of established mathematical models for the spatial spread of
epidemics, perhaps the most widely studied being diffusion models based on
the Fisher--Kolmogorov--Petrovsky--Piskunov family of equations.  These
models are inadequate to represent the phenomena we are interested in,
because diffusion is an inherently local process, equivalent to a disease
transmission process with contacts that span infinitesimal distances only.
To represent the nonlocal transmission implied by our contact kernel, we
use the model of Mollison~\cite{Mollison72}, which is a spatial but
nonlocal susceptible--infectious (SI) model defined as follows.

Let $P(\vec{u},t)$ be the probability that an individual at
position~$\vec{u}$ has the disease at time~$t$ and let $\alpha(r)$ be the
contact kernel, defined as the probability per unit time and area that an
individual has contact sufficient to contract the disease from another a
distance~$r$ away (although transmission will only actually occur if the
first individual does not have the disease and the second does).  For
simplicity we will assume our population to be uniformly distributed in
two-dimensional space.  Real populations are not uniformly distributed,
being concentrated in population centers like towns and villages, and this
could in principle impact travel patterns and hence the spread of disease,
but we will neglect those effects here.  Then the dynamics of the resulting
epidemic is governed by the integro-differential
equation~\cite{Mollison72},
\begin{equation}
  \frac{\partial}{\partial t} P(\vec{u},t)
  = \bigl[1 - P(\vec{u},t)\bigr]
    \int \alpha(|\vec{u}-\vec{v}|) P(\vec{v},t) \>\dd^2\vec{v}.
\label{eq:model}
\end{equation}
The leading factor on the right represents the probability that an
individual at position~$\vec{u}$ is uninfected at time~$t$ and the integral
represents the probability of becoming newly infected by an individual
somewhere else in the plane.  The equation can be regarded as a
continuous-space approximation to the exact dynamics in a discrete
population on, for example, a square grid, where the integral extends over
the grid area.

Analysis of Eq.~\eqref{eq:model} is more complicated than for diffusive
models because of its nonlocality; standard techniques for partial
differential equations cannot be directly applied.  Fortunately, however, we
do not need to solve the equation in complete generality.  Our goal is only
to determine whether wave-like solutions, of the kind seen in the spread of
the Black Death, are possible, and this question turns out to be a
tractable one.

Consider a potential solution to Eq.~\eqref{eq:model} taking the form of a
traveling wave with velocity~$v$.  Without loss of generality, assume the
wave to be traveling in the positive $x$ direction so that, writing
$\vec{u}=(x,y)$, we have $P(\vec{u},t) = f(x-vt)$ for some function~$f$ in
the range $0 \le f \le 1$.  Substituting into Eq.~\eqref{eq:model}, we
obtain \begin{equation} -\!v\frac{\dd f}{\dd x} = \bigl[1-f(x)\bigr]
  \int_{-\infty}^{\infty}\, \beta(x'-x) f(x') \, \dd x',
\label{eq:f}
\end{equation}
where
\begin{equation}
\beta(x) = \int_{-\infty}^\infty \alpha\bigl(\sqrt{x^2+y^2}\bigl) \>\dd y.
\label{eq:beta}
\end{equation}
From the properties of the contact kernel discussed earlier, $\beta$~is
nonnegative, integrable, symmetric about the origin, and nonincreasing for
$x>R$.

The mathematical question we wish to answer is this: what is implied about
$\alpha$ if a solution $0 < f(x) < 1$ to Eq.~\eqref{eq:f} exists?  In
plain terms, if we encounter an epidemic with a stable, steadily advancing
wave-front, what does that tell us about the underlying contact kernel?
Mollison~\cite{Mollison72} showed that if a solution to the equation
exists, then the integral of the tail of~$\beta$ decays exponentially or
faster.  Our first theorem is similar in spirit to this result, but
different in detail and with a more straightforward proof.

\bigskip\noindent\textbf{Theorem~1:} If Eq.~\eqref{eq:f} has a solution $0
< f(x) < 1$, then $\alpha(r) = \Ord(\e^{-cr})$ in the limit $r\to\infty$,
where $c = (1/\sqrt{2}v) \int_0^\infty \beta(x) \>\dd x$.

\bigskip\noindent\textbf{Proof:} Equation~\eqref{eq:f} and the condition
that $0<f<1$ imply that $f$ is continuous (since $f'(x)$ is bounded) and
decreasing (since $f'(x)<0$).

i) Using these facts, we can compute an upper bound on $f(x)$ for $x > 0$.
We note that the integral in Eq.~\eqref{eq:f} only decreases (or stays the
same) if we restrict the domain of integration to the interval
$(-\infty,x)$ and then replace $f(x')$ with~$f(x)$ (because $f(x)$ is
decreasing in~$x$).  Making these changes and then changing the integration
variable to $z = x'-x$, we find
\begin{equation}
-v\frac{\dd f}{\dd x} \ge \bigl[1-f(x)\bigr] f(x) \int_0^\infty
                          \beta(z) \>\dd z,
\label{eq:thm1-1}
\end{equation}
where we have made use of the fact that $\beta$ is symmetric about the
origin.  Equation~\eqref{eq:f} is translation invariant, so without loss of
generality we will let $f(0)=\frac12$.  Separating variables
in~\eqref{eq:thm1-1} and integrating from $0$ to $x$ then gives $f(x) \le
(1+\e^{b x})^{-1}$ for $x \ge 0$, where $b = v^{-1} \int_0^\infty
\beta(z) \>\dd z$.  Thus $f(x)$ obeys the strict inequality
\begin{equation}
f(x) < \e^{-b x} \quad \text{for $x>0$.}
\label{eq:thm1-2}
\end{equation}
In other words, the leading edge of the epidemic decays exponentially or faster.

ii) We now use this result to place bounds on the behavior of~$\beta(x)$
for large~$x$.  To do this we divide Eq.~\eqref{eq:f} by $1-f(x)$ and
integrate from $x$ to $\infty$ to obtain
\begin{equation}
-v \ln\bigl[1-f(x)\bigr] = \int_x^\infty
  \int_{-\infty}^\infty \beta(z-x') f(z) \>\dd z \>\dd x'.
\label{eq:thm1-3}
\end{equation}
Performing a Taylor expansion of the left-hand side about $f=0$ and
applying \eqref{eq:thm1-2}, we arrive at the asymptotic upper bound
\begin{equation}
-v \ln\bigl[1-f(x)\bigr] \lesssim v \e^{-b x} \quad \text{as $x\to\infty$},
\label{eq:thm1-4}
\end{equation}
where $g(x) \lesssim h(x)$ means that for any $\epsilon > 0$ there is an $X
< \infty$ such that $g(x) < (1+\epsilon)h(x)$ for all $x > X$.

Now we restrict the interior integral on the right-hand side
of~\eqref{eq:thm1-3} to the interval $(-\infty,0)$ and replace $f(z)$ with
its smallest value in that interval, which is $f(0)=\frac12$.  Making a
change of variables in the inner integral and noting that $\beta$ is even,
we obtain the inequality
\begin{equation}
\int_x^\infty \int_{-\infty}^\infty \beta(z-x') f(z) \>\dd z \>\dd x'
\ge \tfrac12 \int_x^\infty \int_{x'}^\infty \beta(z) \>\dd z\>\dd x'.
\label{eq:thm1-5}
\end{equation}
Then \eqref{eq:thm1-3}, \eqref{eq:thm1-4}, and~\eqref{eq:thm1-5} together
imply that
\begin{equation}
\int_x^\infty \int_{x'}^\infty \beta(z) \>\dd z \>\dd x' \lesssim 2v \e^{-b x}.
\label{eq:thm1-6}
\end{equation}

iii) This is not yet quite the result we need.  It gives us a bound on a
double integral of~$\beta$.  Our next step gives a bound on $\beta(x)$
itself.

Let $\lambda$ be a positive constant.  As we have said, $\beta(x)$~is
nonincreasing for $x > R$, so the rectangle of height $\beta(x+\lambda)$ from
any $x>R$ to $x+\lambda$ lies under the nonincreasing curve of~$\beta(x)$.
Hence, $\lambda \beta(x+\lambda) \le \int_x^\infty \beta(x') \>\dd x'$
and, setting $\lambda=1/b$, we have
\begin{equation}
\frac{1}{b} \beta(x+1/b) \le \int_x^\infty \beta(x') \>\dd x'.
\label{eq:thm1-7}
\end{equation}
Noting that $\int_x^\infty \beta(x') \>\dd x'$ is also a nonincreasing
function of $x$ for all~$x$ (since $\beta(x)$ is nonnegative everywhere),
we can repeat the same argument again to show that
\begin{equation}
\frac{1}{b} \int_{x+1/b}^\infty \beta(x') \>\dd x'
  \le \int_x^\infty \int_{x'}^\infty \beta(z) \>\dd z \>\dd x'.
\label{eq:thm1-8}
\end{equation}
Then making the substitution $x\to x+1/b$ in \eqref{eq:thm1-7} and combining
with~\eqref{eq:thm1-8} we have
\begin{equation}
\frac{1}{b^2} \beta(x+2/b) \le \int_x^\infty \int_{x'}^\infty
              \beta(z) \>\dd z \>\dd x'.
\label{eq:thm1-9}
\end{equation}
Along with \eqref{eq:thm1-6} this now yields an asymptotic upper bound
on~$\beta(x)$ itself:
\begin{equation}
\beta(x) \lesssim 2vb^2\e^2\,\e^{-bx}.
\label{eq:thm1-10}
\end{equation}

iv) The final step in our proof converts this upper bound on $\beta(x)$ to
an upper bound on the contact kernel~$\alpha(r)$.  From \eqref{eq:thm1-10}
and the definition of~$\beta(x)$, Eq.~\eqref{eq:beta}, we have
\begin{equation}
\int_{-\infty}^\infty \alpha\bigl(\sqrt{x^2+y^2}\bigr) \>\dd y
       \lesssim 2vb^2\e^2\,\e^{-bx}.
\label{eq:thm1-11}
\end{equation}
Integrating both sides with respect to $x$ from an arbitrary $r > R$
to~$\infty$, we obtain
\begin{equation}
\int_r^\infty \int_{-\infty}^\infty \alpha\bigl(\sqrt{x^2+y^2}\bigr)
       \>\dd y \>\dd x \lesssim 2vb\e^2\,\e^{-br}.
\label{eq:thm1-12}
\end{equation}
(Note that the definition of $\lesssim$ permits this.)  The domain of the
double integral in this expression is the region $D = \{ (x,y) \, | \, r\le
x<\infty, -\infty<y<\infty \}$.  The union of four copies of~$D$, rotated
about the origin by $0$, $\pi/2$, $\pi$, and $3\pi/2$, contains the polar
region $P = \{(r',\theta) \, | \, \sqrt{2}r \le r' < \infty, 0 \le \theta <
2\pi \}$.  So if we integrate $\alpha$ over these four copies of $D$ and
sum the results, the sum will be as large or larger than the integral of
$\alpha$ over~$P$:
\begin{equation}
4 \int_r^\infty \int_{-\infty}^\infty
   \alpha\bigl(\sqrt{x^2+y^2}\bigr) \>\dd y \>\dd x \ge
   2\pi \int_{\sqrt{2}r}^\infty \alpha(r')\, r' \>\dd r'.
\label{eq:thm1-13}
\end{equation}
Observing that $r'\alpha(r') > R\alpha(r')$ for all $r'>R$ and that
$\alpha(r')$ is nonincreasing in this regime, we now integrate from
$\sqrt{2}r$ to $\infty$ and apply the same argument as
for~\eqref{eq:thm1-7} to obtain
\begin{equation}
\int_{\sqrt{2}r}^\infty \! \alpha(r')\, r' \>\dd r' >
\frac{\sqrt{2}R}{b} \alpha\bigl(\sqrt{2}r+\sqrt{2}/b\bigr).
\label{eq:thm1-14}
\end{equation}
Then combining \eqref{eq:thm1-12}, \eqref{eq:thm1-13},
and~\eqref{eq:thm1-14} we have
\begin{equation}
\alpha(r) \lesssim
\frac{4\sqrt{2}vc^2\e^3}{\pi R} \, \e^{-cr},
\end{equation}
where $c = b/\sqrt{2} = (1/\sqrt{2}v) \int_0^\infty \beta(x) \>\dd x$.
This completes the proof. $\square$

\medskip This result helps explain the behavior we observed in
Fig.~\ref{fig:TwoSimulations}.  In the first panel we observe distinct,
wave-like spread of the disease from the initial point of infection, and
under these circumstances Theorem~1 tells us that the underlying infection
kernel must have exponential (or faster) decay at long distances, which is
correct in this case.  Conversely, the theorem tells us that if the kernel
decays slower than an exponential at long distances, then we will not
observe wave-like spread, which agrees with the results shown in the second
panel of the figure.

\section{Rare long-range contacts imply no small-world effect}
Given that wave-like behavior implies a contact kernel decaying at least as
fast as an exponential, what now can we conclude about the shape of the
network of disease-causing contacts?  We show in this section that the
network must be a ``large world,'' by which we mean that it does not show
the small-world effect.  The most widely accepted formal definition of the
small-world effect is that typical path lengths in a network increase with
the number of nodes~$n$ no faster than $\log n$~\cite{Newman02,Chung02},
and we will adopt that definition here.  Studies, both theoretical and
empirical, suggest that many networks satisfy this definition and hence
display the small-world effect, but we will show that an exponential
contact kernel produces path lengths that increase significantly faster,
roughly as~$\sqrt{n}/\log n$.

It is important to emphasize that the small-world effect is a mathematical
statement about scaling relationships between path lengths and population
size, not a statement about how path lengths have changed in practice as
the population of the world has grown.  The population of the Earth has
increased during most of human history and that increase has been
accompanied by increases in population density, changes in living
conditions and social customs, and improvements in technology, including
technology for travel and communication.  All of these factors could affect
the shape of social networks and hence might potentially change path
lengths between individuals.  But it is not this process that the
small-world effect describes.  The small-world effect says that even if all
other factors were kept fixed (such as population density, social customs,
and technology), typical path lengths between individuals in the social
network would still increase at most logarithmically with population.  This
is the sense in which we use the phrase.

To demonstrate our results we first need to define exactly what network we
are considering.  We have so far avoided direct reference to the contact
network by working in the continuous space of Eq.~\eqref{eq:model}, which,
as we have said, can be thought of as a continuous approximation to the
disease dynamics of a discrete population.  To address the structure of the
network, we return to the discrete view and consider a population of $n$
individuals uniformly distributed in two-dimensional space.  For
simplicity, let us choose $n$ to be a perfect square and place our $n$
individuals at the sites of an $\sqrt{n}\times\sqrt{n}$ regular square
lattice with unit lattice spacing (though the structure of our argument is
not sensitive to the exact choice of positions for the population members).

We then place edges between individuals in the population to represent the
physical contacts by which disease is or could be transmitted.  We place a
total of exactly $\gamma n$ edges, where $2\gamma$ is the mean degree of
the network---the mean number of connections an individual has.  In keeping
with our assumption above that all social and other parameters are held
constant, we will assume that the mean degree remains constant as $n$ is
varied.  The position of each of the $\gamma n$ edges is drawn
independently from the same probability distribution
$p_n(\vec{u},\vec{v})$, meaning that the ends of an edge fall at positions
$\vec{u}$ and $\vec{v}$ with probability~$p_n(\vec{u},\vec{v})$.  The form
of $p_n(\vec{u},\vec{v})$ depends on~$n$, but if the edges represent
physical contacts then it must assume the same functional form as the
contact kernel~$\alpha(|\vec{v}-\vec{u}|)$ in the limit of large system
size.  Since $p_n(\vec{u},\vec{v})$ must be normalized so that
$\sum_{\vec{u},\vec{v}} p_n(\vec{u},\vec{v}) = 1$, we then have
\begin{equation}
p_n(\vec{u},\vec{v}) \le \frac{\kappa}{n} \alpha(|\vec{v}-\vec{u}|),
\label{eq:pn}
\end{equation}
for some constant~$\kappa$ and sufficient large $n$, given that the contact
kernel is integrable.

We also place additional edges between all pairs of nearest neighbors on
the lattice, which ensures that a path exists between every pair of lattice
sites and simplifies our proof.  These edges need not represent real
physical interactions and do not qualitatively affect our final
result---clearly their addition can only reduce path lengths in the
network, not increase them, so if there is no small-world effect when they
are included then there can be no small-world effect without them either.
Similar nearest-neighbor edges have been used in other models of the
small-world effect~\cite{Kleinberg00,NW99b}, and models of this kind,
consisting of a lattice of nearest-neighbor connections plus random
longer-range ones, are generically known as ``small-world
models''~\cite{Watts98}.

Now consider a pair of sites on the lattice separated by geographic
distance of order the linear dimension~$\sqrt{n}$, which we will write
as~$q\sqrt{n}$ for some constant~$q > 0$.  For instance we might choose a
pair of sites a distance $\frac12\sqrt{n}$ apart.  (Without this
requirement one could trivially find a short path between sites by choosing
sites that just happened to be very close together.)  Let $L_n$ be the
number of hops in the shortest network path between these sites in a
particular realization of the network and let $\E[L_n]$ be the expected
value of~$L_n$ over all realizations.

\bigskip\noindent\textbf{Theorem 2:} If $\alpha(r) = \Ord(\e^{-cr})$ in the
limit $r\to\infty$ for some constant~$c$, then $\E[L_n] =
\Omega(\sqrt{n}/\log n)$ in the limit $n\to\infty$.  Informally, the
expected distance between sites is at least a constant times $\sqrt{n}/\log
n$ in the limit.

\bigskip\noindent\textbf{Proof:} Let $\mathcal{E}$ be the set of edges in a
particular realization of our model and let $(\vec{u},\vec{v}) \in
\mathcal{E}$ indicate that there is an edge between the sites at positions
$\vec{u}$ and~$\vec{v}$.  On a finite lattice there is necessarily,
somewhere in the network, an edge or edges that span the largest geographic
distance, and hence for any length~$\ell(n)\ge1$ that we choose there is a
well-defined probability~$\Pr[ \forall\,(\vec{u},\vec{v})~\in~\mathcal{E} :
|\vec{u}-\vec{v}|\le\ell(n)]$ that all edges will be less than or equal
to~$\ell(n)$ in length.  Moreover, if no edge is longer than~$\ell(n)$,
then the number of hops~$L_n$ in the shortest network path between our two
sites of interest is at least as great as for a path connecting the same
two sites and composed entirely of hops of length~$\ell(n)$.  In other
words $L_n\ge q\sqrt{n}/\ell(n)$.  Combining these observations, it follows
that
\begin{equation}
\Pr[ L_n\ge q\sqrt{n}/\ell(n) ] \ge
  \Pr[ \forall\,(\vec{u},\vec{v})\in\mathcal{E}:|\vec{u}-\vec{v}|\le\ell(n)].
\label{eq:thm2-1}
\end{equation}
Applying Markov's inequality to the left-hand side we have
\begin{equation}
\Pr[ L_n\ge q\sqrt{n}/\ell(n) ] \le \frac{\ell(n)}{q\sqrt{n}} \,\E[L_n],
\end{equation}
and combining this result with~\eqref{eq:thm2-1} gives
\begin{equation}
\E[L_n] \ge \frac{q\sqrt{n}}{\ell(n)} 
\Pr[ \forall\,(\vec{u},\vec{v})\in\mathcal{E}:|\vec{u}-\vec{v}|\le\ell(n)].
\label{eq:thm2-2}
\end{equation}

The function~$\ell(n)$ is undetermined in this expression.  We now show
that if we choose $\ell(n)$ proportional to~$\log n$ then the probability
on the right-hand side of \eqref{eq:thm2-2} is nonzero in the limit
$n\to\infty$, and hence that $\E[L_n]$ grows at least as fast as
$\sqrt{n}/\log n$.  To see this, let $\mathcal{B}_n$ be the set of all
unordered pairs of nodes $(\vec{u},\vec{v})$ such that $|\vec{u} - \vec{v}|
> \ell(n)$.  Then the probability that none of the edges in the network is
longer than~$\ell(n)$ is equal to the probability that none of these pairs
is connected by an edge, which is
\begin{equation}
\Pr[ \forall\,(\vec{u},\vec{v})\in\mathcal{E}:|\vec{u}-\vec{v}|\le\ell(n)]
  = \biggl[1 - \hspace{-1ex} \sum_{(\vec{u},\vec{v}) \in \mathcal{B}_n}\!\!
    p_n(\vec{u},\vec{v}) \biggr]^{\gamma n}.
\label{eq:thm2-3}
\end{equation}
We now note the following points about the contact kernel:
\smallskip
{\renewcommand{\labelenumi}{\roman{enumi})\ }
\begin{enumerate}
\item If $\lim_{n\to\infty} \ell(n) = \infty$, and recalling that the tail
  of $\alpha(r)$ is nonincreasing, there is some $n_1$ such that if $n>n_1$
  then for every $(\vec{u},\vec{v}) \in \mathcal{B}_n$ we have
  $\alpha(|\vec{u}-\vec{v}|) \le \alpha(\ell(n))$.
\item If $\lim_{n\to\infty} \ell(n) = \infty$ and given that $\alpha(r) =
  \Ord(\e^{-cr})$, there is some positive constant $a$ and some $n_2$ such
  that if $n>n_2$ then $\alpha(\ell(n)) \le a\e^{-c\ell(n)}$.
\end{enumerate}}
\smallskip

So if we assume that $\lim_{n\to\infty} \ell(n) = \infty$ and string
together the inequalities in (i), (ii), and \eqref{eq:pn}, we have, for
sufficiently large~$n$,
\begin{equation}
p_n(\vec{u},\vec{v}) \le \frac{\kappa}{n}\alpha(|\vec{u}-\vec{v}|)
  \le \frac{\kappa}{n}\alpha(\ell(n)) \le \frac{\kappa a}{n} \e^{-c\ell(n)}.
\end{equation}
Combining this with Eq.~\eqref{eq:thm2-3} and noting that there are only
$n$ possible $\vec{u}$ sites and $n$ possible $\vec{v}$ sites, so there are
at most $n^2$ terms in the sum in~\eqref{eq:thm2-3}, we have
\begin{equation}
\Pr[ \forall\,(\vec{u},\vec{v})\in\mathcal{E}:|\vec{u}-\vec{v}|\le\ell(n)]
  \ge \bigl[ 1 - n\kappa a \e^{-c\ell(n)} \bigr]^{\gamma n}.
\label{eq:thm2-4}
\end{equation}
Now let $\ell(n) = (2/c) \log n$, which satisfies our requirement that
$\lim_{n\to\infty} \ell(n) = \infty$.  Then the factor $\e^{-c\ell(n)}$
in~\eqref{eq:thm2-4} is just~$1/n^2$.  Combining this observation with
\eqref{eq:thm2-2}, we arrive at
\begin{equation}
\E[L_n] \ge \tfrac12 qc \biggl(1 - \frac{\kappa a}{n} \biggr)^{\gamma n}
\frac{\sqrt{n}}{\log n}.
\label{eq:thm2-5}
\end{equation}
In the limit of large~$n$, we have $(1 - \kappa a/n)^{\gamma n} \to
\e^{-\gamma\kappa a}$ and hence $\E[L_n] = \Omega(\sqrt{n}/\log n)$ in
this limit.  $\square$

\medskip Thus the network within our model, which consists of the network
of physical disease transmission contacts plus the lattice of
nearest-neighbor edges, does not display the small-world effect, and this
immediately implies that the physical contact network alone also does not
display the small-world effect.  As described earlier, the final step in
our argument is to point out, under the assumption that the edges in the
network of social contacts are a subset of the edges in the network of
physical contacts, that the social network will also not display the
small-world effect, since removing edges from the network can only make
paths longer, not shorter.

\section{Discussion}
Using a combination of empirical observation and mathematical reasoning, we
have in this paper argued that the small-world effect---the occurrence of
logarithmically short paths between most individuals in social
networks---is a modern phenomenon, dating back no more than a few hundred
years.  In particular, the observation of slow, wave-like spread in
historical disease outbreaks such as the Black Death strongly suggests that
the social world used to be large.

A number of further points seem worth noting.  First, our calculations are
all performed using an SI model in which infected individuals neither die
nor recover from disease.  However, with most human diseases, the Black
Death included, individuals do recover or die, which could change the
pattern of infection in various ways.  In particular, when infected
individuals remain infectious only for a limited time, the contact kernel
for the disease is effectively truncated at a length-scale corresponding to
the distance a person can travel in that time.  Traveling on horseback, for
example (the fastest form of land transportation in the 14th century), an
individual carrying the Black Death could travel at most about 160 km in
the four days for which victims of the disease remained infectious.  In
effect, therefore, the contact network would have no (or few) connections
beyond this range.  It seems unlikely, however, that this truncation
affects our results.  The Black Death advanced far slower than 40 km per
day---its average speed of progress was perhaps about 2 km a day.  Even on
foot a human can travel an order of magnitude faster than this.  This
suggests that travel velocity was not the limiting factor in the spread of
the disease.  Another simplification in our calculation is the assumption
of a uniform population.  In reality, populations both then and now are
highly nonuniform, being concentrated in metropolitan areas and sparse in
rural areas.  It would make an interesting topic for future study to
incorporate realistically nonuniform distributions into the model, although
it seems likely that one would then no longer be able to derive rigorous
results.

We conclude with a question: if the global social network displays the
small-world effect now but did not in the 14th century, when did things
change?  The occurrence of rapidly spreading pandemics in the 19th century
suggests that the most substantial shift may have been the emergence around
that time of relatively inexpensive means of long-range transportation,
such as commercial railroads and passenger liners, but a full answer to the
question will require detailed historical, geographical, and
epidemiological studies before our understanding is complete.

\begin{acknowledgments}
  We thank Werner Horsthemke for helpful pointers to the literature on
  nonlocal reaction-diffusion equations.  This work was supported by the US
  National Science Foundation under grants DMS--1107796 (TM and MEJN) and
  DMS--0927587 and PHY--1205219 (CRD), by the Marine Alliance for Science
  and Technology for Scotland under grant HR09011 (DL), and by the Michigan
  Society of Fellows (SAM).
\end{acknowledgments}

\end{document}